\documentclass[preprint,showpacs,preprintnumbers,amsmath,amssymb]{revtex4}

\usepackage{graphicx}
\usepackage{dcolumn}
\usepackage{bm}

\begin{document}

\preprint{APS/123-QED}

\title{Ab-Initio Simulations of Deformation Potentials and Electron Mobility in Chemically Modified Graphene and two-dimensional hexagonal Boron-Nitride}

\author{Samantha Bruzzone}
\affiliation{Dipartimento di Ingegneria dell'Informazione: Elettronica, Informatica, Telecomunicazioni, \\Universit\`a di Pisa, Via Caruso 16, 56122 Pisa, Italy.}
\author{Gianluca Fiori}
\affiliation{Dipartimento di Ingegneria dell'Informazione: Elettronica, Informatica, Telecomunicazioni, \\Universit\`a di Pisa, Via Caruso 16, 56122 Pisa, Italy.}

\begin{abstract}

We present an ab-initio study of electron mobility and electron-phonon coupling
in chemically modified graphene, considering fluorinated and
hydrogenated graphene at different percentage coverage. Hexagonal Boron Carbon Nitrogen (h-BCN)
is also investigated due the increased interest shown by the research community
towards this material. 
In particular, the Deformation Potentials are computed by means of
Density Functional Theory (DFT), 
while the carrier mobility is obtained according to the Takagi model 
(S. Takagi, A. Toriumi, and H. Tango, IEEE Trans. Electr. Dev. $\mathbf{41}$, 2363 (1994)).
We will show that graphene with a reduced degree of hydrogenation can compete,
in terms of mobility, with silicon technology.

\end{abstract}

\maketitle

\newpage

Understanding electron mobility in chemically modified graphene is a matter
of primary importance, in order to clarify if graphene will manage to come after
industry requirements in the mid-term.

Despite two dimensional graphene is now at the core of both fundamental and applied research,
 it presents lights and shadows, like its outstanding electronic properties, such as its extremely 
high mobility~\cite{Castro-Neto09,Bolotin08}, but, at the same
time, the lack of an energy bandgap, which prevents its use in digital applications.

Chemical modification of graphene has recently entered the research
scenario as a possible candidate to tackle such a problem.
Elias et al.~\cite{Elias09} managed to experimentally 
induce an energy gap in hydrogenated graphene, 
previously predicted by Sofo et al.~\cite{Sofo07} through numerical calculations, while 
graphene fluorination~\cite{Luo10} has shown a clear 
dependence of the band gap on fluorine concentration.

An alternative option to obtain two-dimensional materials with tunable electronic properties 
has been recently demonstrated by Ci et al.~\cite{Ci10}, exploiting hexagonal Boron Nitride (h-BN) and
 including different percentages of C atoms (h-BCN), or 
$\left(BN\right)_{n}\left(C_{2}\right)_{m}$ $\left(n,m=1,2 \ldots \right)$ isomers. 

However, fabrication techniques are still at an embryonic stage~\cite{Elias09}, so that an experimental
investigation of the main figure of merit for electronic application like electron mobility 
is still far from being performed. 
From this point of view computational approach is the only tool able to provide relevant information regarding 
the intrinsic mobility to be expected. 

To this purpose, in this work we focus on the electron-phonon interaction, in order to
investigate the ultimate intrinsic mobility $\mu$ in
hydrogenated/fluorinated graphene at different percentage 
coverage, as well as in h-BCN structures.

We follow the Bardeen and Shockley's~\cite{Bardeen50} approach, 
where the atomic displacement associated with a long-wavelength acoustic phonon leads
to a deformation of the crystal~\cite{Herring56}, and in turn,  
to a shift of the electronic energy dispersion. 
The band edge shift is related to the differential displacement of the crystal by the 
electron-phonon coupling Hamiltonian~\cite{Long09,Murphy10,Xu10}:

\begin{equation}
H_{el-ph} = E_1 \nabla \cdot \vec u \left(\vec r,t\right) 
\label{DP1}
\end{equation}

where $E_1$ is the so-called deformation potential and $\vec u\left(\vec r,t\right)$ is the displacement at the spatial coordinate $\vec r$~\cite{STN}.
In order to evaluate $\nabla \cdot \vec u\left(\vec r,t\right)$, we define a second-rank tensor 
$d_{ij} = \frac{\partial u_i }{\partial x_j}=e_{ij}+f_{ij}$,
where $u_i$ and $x_i$ are the $i-th$ component of vector $\vec u$ and the cartesian coordinates, respectively, 
while $e_{ij}$ and $f_{ij}$ are the antisymmetric and symmetric tensors, respectively.
In particular, $e_{ij}$ describes a rigid rotation of the crystal and does not lead to a shift of the electronic energies,
so that, to our purposes, it can be ignored.
$f_{ij}$ instead describes a strain induced in the crystal by the atomic displacements, which indeed is able to  
shift the bands. By definition, its trace is equal to the 
fractional volume change $\frac{\delta V}{V_0}$ of the elementary cell induced by the strain.
Representing the long wavelength acoustic vibration by a plane wave with frequency $\omega$ 
and wave vector $\vec q$ as $\vec u\left(\vec r,t \right) = \delta u_0 \sin\left( \vec q \cdot \vec r - \omega t  \right)$,
the strain tensor $f_{ij}$ in proximity of the $\Gamma$ point, for the longitudinal acoustic (LA) mode,  
is diagonal, i.e. $f_{ii} = q_i\delta u_{0i}$,
where $q_i$ and $\delta u_{0i}$ are the $i-th$ component of the wave vector 
$\vec q$ and the atomic displacement from equilibrium geometry, respectively.  
In this  hypothesis, since $\vec q \cdot \vec u = \frac{\delta V}{V_0}$,
we obtain $\nabla \cdot \vec u = \frac{\delta V}{V_0}.$

For non-degenerate bands~\cite{deWalle88}, Eq.(\ref{DP1}) can be eventually reduced to

\begin{equation}
\delta E\left(k\right) = E_1 \left( \frac{\delta V}{V_0} \right)
\label{DP2}
\end{equation}

where $\delta E\left( k \right)$ is the induced band edge shift due to acoustic phonon.
Once known $E_1$, the carrier mobility can be estimated by means of the Takagi formula~\cite{Takagi94}, which reads

\begin{equation}
\mu = \frac{e \hbar^3 \rho S_{l}^2}{k_{B} T m_{e} m_{d} E_{1}^2}
\end{equation}

where $\rho$ is the mass density, $T$ is the temperature, $k_B$ the Boltzmann's constant, $e$ is the elementary charge,
$S_l$ is the sound velocity, 
while $m_e$ is the mass along the transport direction (either $m_x$ or $m_y$ along the 
$x$ and $y$ direction, respectively) and $m_d$ is the equivalent density-of-state
 mass defined as $m_d=\sqrt{m_xm_y}$.
$S_l$, $m_e$ and $m_d$ are computed by means of ab-initio calculation, and are extracted from the 
phonon $\omega(k)$ and electronic dispersion relations $E(k)$,
according to $S_{l}=\left[\frac{\partial \omega \left(k\right)}{\partial k}\right]$ and
$m_{e}=\hbar\left[\frac{\partial^{2}E\left(k\right)}{\partial k^2}\right] ^{-1}$. 
The deformation potential $E_1$ is instead computed through mimicking the lattice deformation due to phonons by 
multiplying the lattice constant by different factors (0.99, 0.995, 1.005 and 1.01), i.e. dilating 
and relaxing the primitive cell. 
$E_1$ is then obtained computing the energy shift of the bottom of the conduction band $\delta E$ 
and applying Eq.(\ref{DP2}).

Ab-initio calculations have been performed by means of the Quantum Espresso~\cite{QE} code, using a plane 
wave basis set in the generalized gradient approximation (GGA) with the Perdew-Burke-Ernzerhof (PBE) 
exchange correlation functional. 
A 50 Ry wave function cutoff and 400 Ry charge density cutoff have been considered, while
the Brillouin zone has been sampled using a $30 \times 30 \times 2$ 
Monkhorst$-$Pack grid. A 30 bohr layer of vacuum is considered to separate the sheet from its periodical 
images, which we have verified to be sufficient to avoid any unphysical interactions. The dynamical 
properties are calculated within the density-functional perturbation theory~\cite{Baroni01}. 
We have also verified that extracting the main parameter of interest, i.e. the effective
mass, from DFT-GGA simulations manages to provide results in close agreements with those
extracted from GW calculations (within 5\% of accuracy), but with
fewer computational requirements (an overall simulation walltime almost 3 order of magnitude smaller).

\begin{figure} [tbp]
\vspace{0.4cm}
\begin{center}
\includegraphics[width=12cm]{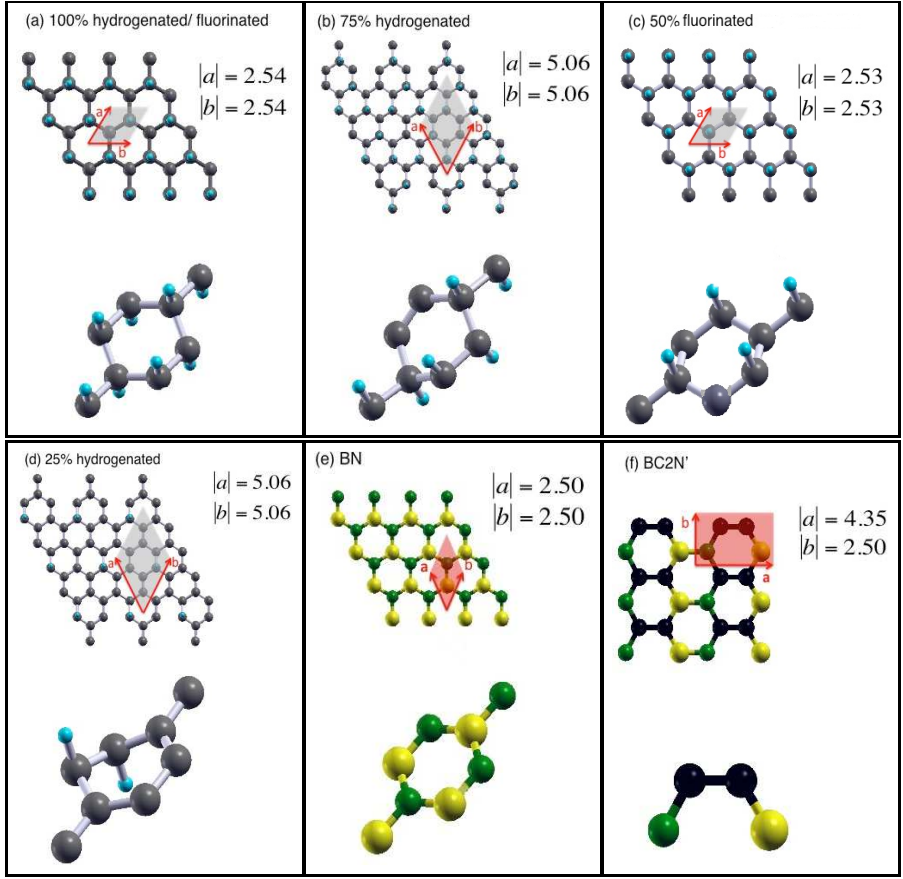}
\end{center}
\caption{Geometric structures of the considered 2D materials: graphene sheet with
 different hydrogen and fluorine coverage (a) 100$\%$; (b) 75$\%$; (c) 50$\%$; (d) 25$\%$;
(e) BN and (f) BC$_2$N. The primitive cell and the lattice vectors are also shown for each material.}
\label{geom}
\end{figure}

In Fig.~\ref{geom}, we show the considered structures. 
Relaxed structures for 100\% hydrogenated and fluorinated graphene and 50\% fluorinated graphene (from here on H100\%, 
F100\% and F50\%, respectively)
are in agreement with results already shown in the literature ~\cite{Sofo07,Sahin11}. 
25\% and 75\% hydrogenated graphene (H25\% and H75\%) 
have been studied considering a cell composed by 8 C atoms and 2 H  
and 6 H, respectively.
In particular, the topological shown in~Fig.\ref{geom} is the most energetically favourable over all  
possible geometric and spin configurations.
Simple chemical consideration can support these results. The relaxed H25\% configuration 
is the only one possessing a resonant chain of alternating double and single bonds, 
which guarantees structure stability.
On the other hand, the considered H75\% structure shows a double bond inside the cell, 
which assures a stable structure. All the structures are nonmagnetic, since
 unsaturated C atoms are all neighbors and their $p_{z}$ orbitals form $\pi$-bonding, which quenches
magnetism.

All the considered materials (except F50\%, which is a metal) are semiconducting and have non-degenerate 
conduction band minimum at the $\Gamma$ point. 

\begin{table}
\begin{center}
\caption{\label{table1} Density mass $\rho$, sound velocity $S_l$, relative effective mass $m_e$, deformation potential $E_1$ for 
the considered structures.}
\begin{tabular}{llllll}
\hline
\       &   $\rho$                                      &  $S_l$                                     &effective         & effective         & $E_1$               \\
\       &   $\left(\times 10^{-7}\frac{Kg}{m^2}\right)$ &  $\left(\times 10^{4}\frac{m}{sec}\right)$ &reduced           & reduced           & $\left( eV \right)$ \\ 
\       &                                               &                                            &mass ($\Gamma K$) & mass ($\Gamma M$) &                     \\
\hline
\ H 100\% &  $7.73 $ & $1.78 $ & 1.02 & 1.02 & 6.9   \\
\ H 75\%  &  $7.65 $ & $2.21 $ & 1.59 & 1.39 & 8.58  \\
\ H 25\%  &  $7.35 $ & $2.46 $ & 0.12 & 0.53 & 9.44  \\
\\     
\ F 100\% &  $18.85 $ & $1.18 $ & 0.53 & 0.53 & 20.84 \\  
\ F 50\%  &  $12.98 $ & $1.34 $ & 0.99 & 0.99 & 10.94 \\  
\\     
\ h-BN    &  $7.59  $ & $1.94 $ & 0.97 & 0.97 & 3.66  \\
\         &                        & ($\Gamma X$)\;\;\;\;($\Gamma Y$)          & ($\Gamma X$)& ($\Gamma Y$)& \\     
\ B$C_2$N &  $7.46  $ & $2.1 $;  $1.89 $ & 0.39 & 0.27 & 18.74  \\
\end{tabular}
\end{center}
\end{table}

The effective masses shown in Tab.~\ref{table1}
are evaluated through a five-point second derivative, considering
a k-point spacing smaller than $0.01 \AA^{-1}$ in order to avoid nonparabolic effects.
The sound velocities are extracted from the slope of the acoustic longitudinal phonon branch.
H25\% and H75\% show different values for the effective mass along the 
$\Gamma K$ and $\Gamma M$ directions, leading to two different values of electron mobility.
The relative effective mass is quite large and close to one for the larger degree of 
functionalization. At the lowest degree of H coverage, the electron 
effective mass is small, since in the limit of no functionalization, the null 
mass of graphene has to be recovered. 

In order to understand the deformation potential $E_1$ trend as a function of H percentage,
 we focus on the LUMO (Lowest Unoccupied Molecular Orbital) at the $\Gamma$ point.

\begin{figure} [tbp]
\vspace{0.4cm}
\begin{center}
\includegraphics[width=12cm]{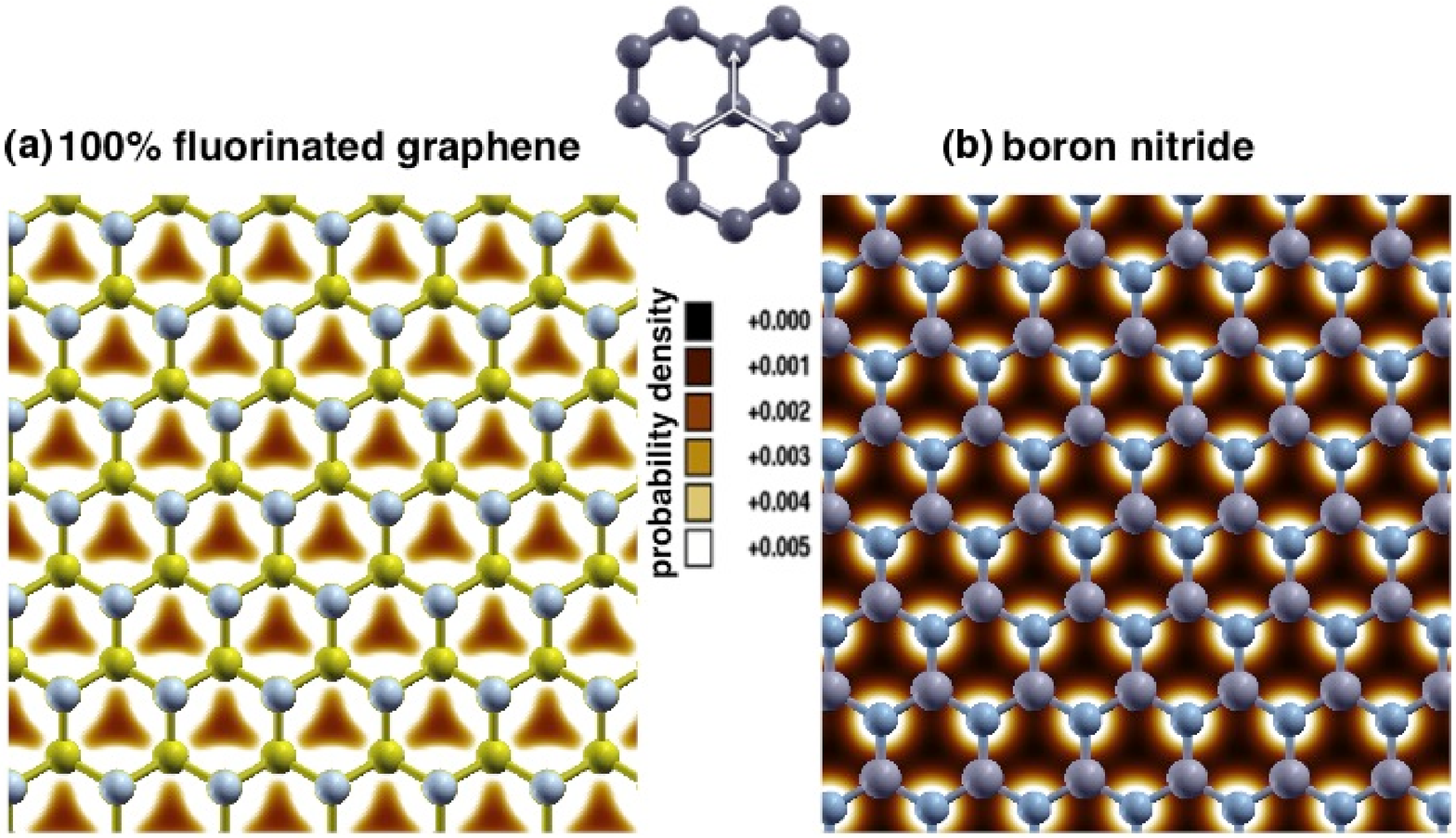}%
\end{center}
\caption{$\Gamma$-point LUMO wave function for (a) 100\% fluorinated and (b) h-BN. In the
picture at the top the direction of the stretching is depicted (white arrows).}
\label{LUMO}
\end{figure}

In Fig.~\ref{LUMO}, we show the LUMO for the F100\% and h-BN, corresponding 
to the largest and smallest obtained $E_1$, respectively
(the lighter the colour, the higher the probability density).
In particular, the h-BN shows the LUMO strongly localized on single atoms (Fig.~\ref{LUMO}b), 
while in the F100\% case the LUMO is spread over groups of atoms (Fig.~\ref{LUMO}a).
Since the deformation of the crystalline lattice due to longitudinal acoustic phonon is represented  
as an homogeneous, symmetrical dilation on the plane 
(picture at the top, where the direction of the stretching is depicted by the arrows in white), the LUMO of F100\% 
is largely deformed by the phonon motion, leading to a stronger electron-phonon coupling, 
and consequently, to large value of $E_1$.
On the other hand, when the LUMO is strongly localized,
the electron-phonon coupling is poor and the deformation potential $E_1$ is consequently low,
as in the case of h-BN.

In Tab.~\ref{table2}, the electron mobilities computed for all the considered materials
along the main symmetry directions are shown. H25\%, H75\% and B$C_2$N 
exhibit different mobilities along the two 
considered directions, as a consequence of the asymmetry of the conduction energy band.

\begin{table}
\begin{center}
\caption{\label{table2} Electronic mobility $\mu$ of the considered materials along the main directions of symmetry.}
\begin{tabular}{lll}
\hline
\       &  $\mu \left(\frac{cm^2}{Vsec}\right)$      &  $\mu \left(\frac{cm^2}{Vsec}\right)$  \\
\       &  $\left( \Gamma K \right) $                &  $\left( \Gamma M \right) $   \\
\hline

\ H 100\% & 104.8    & 104.7   \\
\ H 75\%  &  45.7    &  52.1   \\
\ H 25\%  & 3597.6    & 805.1  \\
\\
\ F 100\% & 44.9     & 44.9   \\
\ F 50\%  & 42.6     & 42.6   \\
\\
\ BN      & 486.9    & 486.9  \\
\       &  $\left( \Gamma X\right) $                &  $\left( \Gamma Y \right) $   \\
\ B$C_2$N & 153.3    & 179.8 \\
\end{tabular}
\end{center}
\end{table}

The low mobility values for highly hydrogenated graphene are in quantitative agreement
with available experimental data recently reported in the literature as a function of the exposure
time to H~\cite{Jaiswall11}. 
While fluorinated compounds show the smallest mobilities among
the considered structures, H25\% possess mobility comparable to bulk silicon,
and an energy gap suitable for electronic applications, which could lead to a possible
exploitation of such material in next-generation nanoscale devices.
Finally, while BC$_2$N shows poor mobility in both $\Gamma X$ and $\Gamma Y$ directions, $\mu$ in
BN is large enough for digital applications.

To conclude, we have computed the electron mobility of hydrogenated and fluorinated graphene 
as well as h-BCN from first-principles. 
The adopted method manages to provide accurate results with reduced computational 
requirements, since it avoids explicit calculation of electron-phonon scattering coefficients. 
Deformation potentials have been extracted from DFT calculations, which
can be useful for future works dealing with Monte Carlo simulations of transport in 
functionalized graphene based transistors. Among the considered structures, H25\% has better performance in terms of mobility, showing $\mu$
comparable with bulk Silicon, which could open its exploitation as device channel in the next technological nodes.
As a word of caution, we have to point out that the obtained mobility represents an upper limit
 for the intrinsic mobility achievable in such structures, since 
additional degree of inhomogeneity (not considered in the present work) could lead to mobility
reduction.

Authors gratefully acknowledge support from the EU FP7 Project 
NANOSIL (n. 216171), GRAND (n. 215752) grants, 
and by the MIUR-PRIN project GRANFET (Prot. 2008S2CLJ9) via 
the IUNET consortium.

\end{document}